\documentclass[3p]{elsarticle}
\biboptions{sort&compress}

\usepackage{amsmath,amssymb}
\usepackage{graphicx}

\newcommand{\be}{\begin{equation}}
\newcommand{\ee}{\end{equation}}
\newcommand{\ben}{\begin{eqnarray}}
\newcommand{\een}{\end{eqnarray}}
\newcommand{\esp}[1]{\left< #1 \right>}
\newcommand{\cor}[1]{\left[ #1 \right]}
\newcommand{\pare}[1]{\left( #1 \right)}
\newcommand{\key}[1]{\left\{ #1 \right\}}
\newcommand{\bc}{\begin{center}}
\newcommand{\ec}{\end{center}}

\begin{document}

\begin{frontmatter}

\title{Direct spreading measures of Laguerre polynomials}

\author[ma,cp]{P. S\'anchez-Moreno}
\ead{pablos@ugr.es}
\author[fm,cp]{D. Manzano}
\ead{manzano@ugr.es}
\author[fm,cp]{J.S. Dehesa}
\ead{dehesa@ugr.es}

\address[ma]{Departamento de Matem\'atica Aplicada, Universidad de Granada, Granada, Spain}
\address[fm]{Departamento de F\'{\i}sica At\'omica, Molecular y Nuclear, Universidad de Granada, Granada, Spain}
\address[cp]{Instituto ``Carlos I'' de F\'{\i}sica Te\'orica y Computacional, Universidad de Granada, Granada, Spain}

\begin{abstract}
The direct spreading measures of the Laguerre polynomials $L_n^{(\alpha)}(x)$, 
which quantify the distribution of its Rakhmanov probability density $\rho_{n,\alpha}(x)=\frac{1}{d_n^2} x^\alpha e^{-x}\left[L_n^{(\alpha)}(x)\right]^2$ along the positive real line in various complementary
and qualitatively different ways, are investigated. These measures include the 
familiar root-mean-square or standard deviation and the information-theoretic
lengths of Fisher, Renyi and Shannon types. The Fisher length is explicitly given. The
Renyi length of order $q$ (such that $2q\in\mathbb{N}$) is also found in terms of $(n,\alpha)$ by means of two 
error-free computing approaches; one makes use of the Lauricella function
$F_A^{(2q+1)}\left( \frac{1}{q},...,\frac{1}{q};1 \right)$, which is based on 
the Srivastava-Niukkanen linearization relation of Laguerre polynomials, 
and another one which utilizes the multivariate Bell polynomials of Combinatorics. 
The Shannon length cannot be exactly calculated because of its logarithmic-functional form, but its asymptotics is provided and sharp bounds are obtained by use of an
information-theoretic optimization procedure. Finally, all these spreading measures 
are mutually compared and computationally analyzed; in particular, it is found 
that the apparent quasi-linear relation between the Shannon length and 
the standard deviation becomes rigorously linear only asymptotically (i.e. for 
$n>>1$).
\end{abstract}


\end{frontmatter}

\section{Introduction}
The Laguerre polynomials $\left\{L_n^{(\alpha)}(x)\right\}$ are real hypergeometric polynomials orthonormal with respect to the weight function $\omega_\alpha(x)=x^\alpha e^{-x}$ on the interval $[-1,1]$.
Their algebraic properties (orthogonality relation,
three-term recurrence relation, second-order differential equation, ladder relation,...) 
are simple and widely  known \cite{chihara_78,nikiforov_88,andrews_99}, what has enabled to describe a great deal of scientific and technological phenomena.
The Laguerre polynomials are known to play a crucial role not only in applied and computational mathematics \cite{chihara_78,andrews_99,ismail_05,temme_96,nikiforov_88,chen:mcm03,srivastava:ass88,hounkonnon:ame00,srivastava_85}, mathematical physics \cite{nikiforov_88,crandall:jcp85}, combinatorics \cite{foata_82,desainte:lnm85,foata:sjdm88,askey_jct78}, information theory \cite{yanez:jmp08,dehesa:jcam01,sanchezmoreno:preprint09,aptekarev:jcam09,dehesa:jcam06,sanchezruiz:jcam05,aydmer_arxiv06}, quantum algebras \cite{srivastava:ass88,dattoli_arxiv07}, asymptotics \cite{aptekarev:jcam09,borwein:sjna08,aptekarev:rassm95,dehesa:jmp98,sanchezruiz:jcam00} and theory of special functions \cite{temme_96,nikiforov_88,yanez:jmp08,dehesa:jcam01,sanchezmoreno:preprint09,aptekarev:jcam09,srivastava:mcm03,dehesa:jcam06,sanchezruiz:jcam05,niukkanen:jpa85,devicente_04}, but also in non-relativisitic, relativistic and supersymmetric quantum mechanics \cite{galindo_pascual_90,bagrov_90,cooper_01,flugge_71,niukkanen:jpa85}, atomic and molecular physics \cite{sanchezruiz:jcam03,bransden_03,dehesa:mp06,aydmer_arxiv06} and D-dimensional physics \cite{wang:fpl02,nieto_ajp79,dong:mpl05}.
Let us just mention that the Laguerre polynomials control to a great extent the wavefunctions which describe the quantum states of one and many-body systems 
with a great diversity of quantum-mechanical potentials \cite{galindo_pascual_90,sanchezmoreno:preprint09,omiste:jmc09,flugge_71,sanchezruiz:jcam03,dunkl:aa03,dehesa:mp06,nieto_ajp79,sanchezruiz:jcam00,aydmer_arxiv06,gangopadhyay_jpa98}.
The Coulomb and Morse potentials are only two particularly relevant cases in atomic
 and molecular physics (see e.g. \cite{bagrov_90,flugge_71,crandall:jcp85,bransden_03,dunkl:aa03})
 as well as in $D$-dimensional physics \cite{wang:fpl02,alves:jpa88}, where the radial wavefunctions 
 are controlled by Laguerre polynomials.

In this work we study the spreading measures of Laguerre polynomials $L_n^{(\alpha)}$, which 
quantify the distribution of its associated probability density
\be\label{eq:density}
\rho_{n,\alpha}(x)=\frac{1}{d_n^2}\left[ L_{n}^{(\alpha)}(x) \right]^2 \omega_\alpha (x),
\ee
where $d_n^2$ is a constant such that $\rho_{n,\alpha}(x)$ is normalized to the unity.
Heretoforth this distribution is called the Rakhmanov density of the polynomials, in honour to the pioneering work 
\cite{rakhmanov:mus77} of this mathematician who has 
shown that this density governs the asymptotic 
$(n\to \infty)$ behaviour of the ratio of two orthogonal polynomials with
contiguous orders $n$ and $n+1$. Physically, this probability density 
characterizes the stationary states of a large class of quantum-mechanical 
potentials \cite{sanchezmoreno:preprint09,bagrov_90,yanez:jmp08,cooper_01,
nikiforov_88,omiste:jmc09}. The most familiar
spreading measure is the simple root-mean-square or standard deviation
\be\label{eq:variance}
\left ( \Delta x \right)_{n,\alpha}=\left(
\left< x^2 \right>_{n,\alpha}-
\left< x \right>^2_{n,\alpha}
\right)^{\frac{1}{2}},
\ee
where the expectation value of a function $f(x)$ is defined by
\be\label{eq:esp}
\esp{ f(x)}_{n,\alpha}:=\int_{0}^{\infty} f(x) \rho_{n,\alpha}(x)dx.
\ee

The information-theoretic-based spreading measures of the Laguerre 
polynomials are not so well known. We refer to the Fisher information
\be\label{eq:fisher}
F\left[ \rho_{n,\alpha} \right]:= \esp{\cor{\frac{d}{dx}\log{\rho_{n,\alpha} (x) } }^2}
= \int \frac{\cor{\rho_{n,\alpha}'(x)}^2}{\rho_{n,\alpha}(x)}dx,
\ee
to the Renyi entropy of order $q$
\be\label{eq:renyi}
R_q\cor{\rho_{n,\alpha}}:= \frac{1}{1-q}\log{\esp{\cor{ \rho_{n,\alpha} (x) }^{q-1}}},
\ee
and its $(q\to1)$ limit, the Shannon entropy
\[
S\cor{\rho_{n,\alpha}}:=-\int_0^\infty  \rho_{n,\alpha} (x)  \log{ \rho_{n,\alpha} (x) } dx,
\]
which measure the distribution of the Laguerre polynomial $L_n^{(\alpha)}$ all over the 
orthogonality interval without reference to any specific point of the interval, so providing
alternative and complementary measures for the spreading of the Laguerre polynomials.
The knowledge of these measures and some quantum-mechanical applications is reviewed in Ref.
 \cite{dehesa:jcam01} up to 2001. Their behaviour for large $n$ and fixed $\alpha$
has been recently surveyed \cite{aptekarev:jcam09}.

Since the Fisher, Renyi and Shannon measures of a given density $\rho(x)$ have
particular units, which are different from that of the variable $x$, it is much more useful
 to use the related information-theoretic lengths 
 \cite{hall:pra99,hall:pra00,hall:pra01,sanchezmoreno:jcam09,guerrero:preprint10}; 
 namely, the Fisher length given by
\be\label{eq:flength}
\pare{\delta x}_{n,\alpha}=\frac{1}{\sqrt{F\cor{\rho_{n,\alpha}}}},
\ee
and the $qth$-order Renyi and Shannon lengths defined by
\be\label{eq:rlength}
{\cal L}_q^R\cor{\rho_{n,\alpha}}=\exp{\pare{R_q\cor{\rho_{n,\alpha}}}};\quad q>0,q\neq1,
\ee
and
\be\label{eq:shannon}
N\cor{\rho_{n,\alpha}}=\lim_{q\to1} {\cal L}_q^R\cor{\rho_{n,\alpha}}=
\exp{\pare{S\cor{\rho_{n,\alpha} }}},
\ee
respectively. Following Hall \cite{hall:pra99,hall:pra00,hall:pra01}, these three quantities together with the standard
deviation will be referred as the direct spreading measures of the density $\rho_{n,\alpha}$
 because they share the following properties: linear scaling under adequate boundary conditions, same units as the variable, and vanishing when the density tends to a delta density.
Moreover, they have an associated uncertainty property \cite{hall:pra99,hall:pra00,hall:pra01,dehesa:mp06} and fulfil the inequalities
\be\label{eq:inequalities}
\pare{\delta x}_{n,\alpha} \le \pare{\Delta x}_{n,\alpha}\qquad
 \textrm{and} \qquad N\cor{\rho_{n,\alpha}}\le 
\pare{ 2\pi \textrm{e}}^\frac{1}{2} \pare{\Delta x}_{n,\alpha},
\qquad \text{when } \rho_{n,\alpha}(0)=0.
\ee

Here we will investigate the direct spreading measures of the Laguerre polynomials
 mentioned above. First, in Section II, we give the known values of the ordinary moments, 
the standard deviation and the Fisher length of these polynomials. Second, in Section III, 
the entropic moments $\esp{\cor{\rho_{n,\alpha}(x)}^k}$ and the Renyi lengths are 
computed by use of two different approaches; one makes use of the Srivastava-Niukkanen 
linearization relation \cite{srivastava:mcm03} of Laguerre polynomials, and another one 
 based on the combinatorial multivariate Bell polynomials \cite{riordan_80,sanchezmoreno:jcam09}.
Third, in Section IV, the asymptotics of the Shannon length is given and some sharp 
bounds to this measure are found. Then, in Section IV, all the four spreading
 measures are computationally discussed. Finally, some open problems and 
conclusions are given.

\section{Ordinary moments, standard deviation and Fisher length}

In this section the known values for the moments-around-the-origin 
$\esp{x^k}_{n,\alpha}$ $(k \in \mathbb{Z} )$, the standard deviation
$\pare{\Delta x}_{n,\alpha}$ and the Fisher length
 $\pare{\delta x}_{n,\alpha}$  of the Laguerre polynomials are given. 
Let us start writing the orthonormality relation 
\[
\int_0^\infty \widetilde{L}_{n}^{(\alpha)} (x) \widetilde{L}_{m}^{(\alpha)} (x)
\omega_\alpha (x)dx=\delta_{nm},
\]
for the orthonormal Laguerre polynomials
\be\label{eq:normallaguerre}
 \tilde{L}_{n}^{(\alpha)} (x)
=\cor{\frac{n!}{\Gamma (\alpha +n+1)}}^{\frac{1}{2}} L_{n}^{(\alpha)} (x).
\ee

Then, the moment-around-the-origin of order $k \in \mathbb{Z}$ is 
defined, according to Eq. (\ref{eq:esp}), by
\[
\esp{x^k}_{n,\alpha}=\int_0^\infty
x^k\rho_{n,\alpha}(x)dx=
\int_0^\infty x^{k+\alpha}e^{-x}
\cor{\widetilde{L}_n^{(\alpha)}(x)}^2dx.
\]
This integral can be calculated by different means; in particular, by use of the expression \cite{nieto_ajp79,schrodinger_ap26_1,schrodinger_ap26_2,schrodinger_ap26_3,schrodinger_ap26_4}
\[
 \int_0^\infty x^s e^{-x} L_n^\alpha (x) L_m^\beta (x) dx = 
\Gamma(s+1) \sum_{r=0}^{{\rm min}(n,m)}
(-1)^{n+m} \binom{s-\alpha}{n-r} \binom{s-\beta}{m-r}\binom{s+r}{r},
\]
one finds that
\[
\esp{x^k}_{n,\alpha}=\frac{n!\Gamma(k+\alpha+1)}{\Gamma(n+\alpha+1)} \sum_{r=0}^n
\binom{k}{n-r}^2\binom{k+\alpha+r}{r},
\]
where the binomial number is 
$\binom{a}{b}=\frac{\Gamma(a+1)}{\Gamma(b+1)\Gamma(a-b+1)}$.
Then, taking into account Eq. (\ref{eq:variance}) and the values $\esp{x^k}_{n,\alpha}$ for 
$k=1,2$, one has the following expression for the standard deviation of 
the Laguerre polynomials \cite{dehesa:jcam06}
\be\label{eq:deltax}
\pare{\Delta x}_{n,\alpha} = \sqrt{2n^2+2(\alpha +1)n+\alpha+1}.
\ee

The Fisher information of the Laguerre polynomials defined by Eq. (\ref{eq:fisher})
 has been recently shown \cite{sanchezruiz:jcam05,yanez:jmp08} to have the value
\begin{eqnarray*}
F\cor{\rho_{n,\alpha}}=
\left\{
\begin{array}{cc}
4n+1;& \alpha=0,\\
\frac{(2n+1)\alpha+1}{\alpha^2-1};& \alpha>1,\\
\infty;&  \alpha\in (-1,+1],\alpha\neq 0,\\
\end{array}
\right.
\end{eqnarray*}
so that the Fisher length of these polynomials has, according to (\ref{eq:flength}),
 the value
\begin{eqnarray*}
\pare{\delta x}_{n,\alpha}=
\left\{
\begin{array}{cc}
\frac{1}{\sqrt{4n+1}}; & \alpha=0,\\[2mm]
\sqrt{\frac{\alpha^2-1}{(2n+1)\alpha+1}};&\alpha>1,\\[2mm]
0;&  \alpha \in (-1,+1],\alpha\neq 0.\\
\end{array}
\right.
\end{eqnarray*}

It is worth remarking that the inequality $(\delta x)_{n,\alpha}\le(\Delta x)_{n,\alpha}$ is 
clearly satisfied.

\section{Renyi lengths}
In this section the Renyi lengths of the Laguerre polynomials 
$\mathcal{L}_q^R \cor{\rho_{n,\alpha}}$ defined by Eq. (\ref{eq:rlength}) will be computed by 
two different approaches: an algebraic approach which is based on the Srivastava-Niukkanen 
linearization relation \cite{srivastava:mcm03}, and a combinatorial method which
utilizes the multivariate Bell polynomials \cite{riordan_80}.
Let us advance that the Renyi integrals are computed only for the half-integer values of $q$. It is worth noting that the final formulas cannot clearly be extended to all real values of $q$, which explains why we cannot take limits
to get the Shannon entropy in a straightforward manner.

According to Eqs. (\ref{eq:renyi}) and (\ref{eq:rlength}), the Renyi length of order 
$q$ is given by
\be\label{eq:lq}
{\cal L}_q^R\cor{\rho_{n,\alpha}}=\key{ W_q\cor{\rho_{n,\alpha}}}^{-\frac{1}{q-1}},\quad q>0,q\ne 1,
\ee
where
\be\label{eq:wq}
W_q\cor{\rho_{n,\alpha}}:=\esp{\cor{\rho_{n,\alpha}(x)}^{q-1}}=\int_0^\infty 
\cor{\rho_{n,\alpha}(x)}^{q}dx=\int_0^\infty \cor{\widetilde{L}_n^\alpha (x)}^{2q}
x^{q\alpha}e^{-qx}dx,
\ee
are the frequency or entropic moments of the Rakhmanov density (\ref{eq:density}) of the
Laguerre polynomials. In spite of the efforts of numerous researchers
\cite{andrews_99,foata:sjdm88,desainte:lnm85,foata_82,niukkanen:jpa85,chen:mcm03,
 srivastava:ass88,hounkonnon:ame00}, these quantities have not yet been calculated. 
Here we will compute them for $2q\in\mathbb{N}$ by use of two different approaches.

\subsection{Algebraic approach}
 To calculate the entropic moment $W_q\cor{\rho_{n,\alpha}}$, we first use 
(\ref{eq:normallaguerre}) and (\ref{eq:wq}) to write
\be\label{eq:wq2}
W_q\cor{\rho_{n,\alpha}}:=\cor{\frac{n!}{\Gamma(\alpha+n+1)}}^q  I_q\cor{L_n^{(\alpha)}},
\ee
where
\be\label{eq:iq}
I_q\cor{L_n^{\alpha}}:=\int_0^\infty x^{\alpha q} e^{-qx} \cor{L_n^{(\alpha)}(x)}^{2q} dx.
\ee

This functional of the orthogonal Laguerre polynomial can be calculated by use of the
linearization formula of Srivastava-Niukkanen \cite{srivastava:mcm03} for the products of 
 various Laguerre polynomials given by
\[
x^\mu L_{m_1}^{(\alpha_1)}(t_1x) \cdots L_{m_r}^{(\alpha_r)}(t_rx)=\sum_{k=0}^\infty 
  \Theta_k \pare{\mu,\beta,r,\{m_i\},\{\alpha_i\},\{t_i\}}L_{k}^{(\beta)}(x),
\]
where the coefficients $\Theta_k \pare{\mu,\beta,r,\{m_i\},\{\alpha_i\},\{t_i\}}$ can be expressed as
\begin{eqnarray*}
&\Theta_k \pare{\mu,\beta,r,\{m_i\},\{\alpha_i\},\{t_i\}}= (\beta+1)_{\mu} \binom{m_1+\alpha_1}{m_1}
\cdots \binom{m_r+\alpha_r}{m_r} \\
&\times F_A^{(r+1)}\cor{\beta+\mu+1,-m_1,\cdots,-m_r,-k;\alpha_1+1,\cdots,\alpha_r+1,\beta+1;
t_1,\cdots,t_r,1},
\end{eqnarray*}
in terms of the Lauricella's hypergeometric functions of $(r+1)$ variables 
\cite{srivastava_85}. The Pochhammer symbol is
 $(a)_n =\frac{\Gamma(a+n)}{\Gamma(a)}$. This general relation 
with the values ($\beta=0$,  $\alpha_1=\cdots=\alpha_r=\alpha$, 
$m_1=\cdots=m_r=n$, $x=qt$, $t_1=\cdots=t_r=\frac{1}{q}$, 
$\mu= \alpha q,r=2q$) readily yields the following linearization result for the powers of
Laguerre polynomials:
\be\label{eq:line}
\pare{qt}^{2q} \cor{L_n^{(\alpha)}(t)}^{2q}=\sum_{k=0}^{\infty}
\Theta_k \pare{\alpha q,0,2q,\{n\},\{\alpha\},\left\{\frac{1}{q}\right\}} L_k^{(0)}(qt),
\ee
where
\begin{multline*}
\Theta_k \pare{\alpha q,0,2q,\{n\},\{\alpha\},\left\{\frac{1}{q}\right\}} = \Gamma(\alpha q+1)\binom{n+\alpha}{n}^{2q}\\ \times F_A^{(2q+1)}\left(\alpha q+1;-n,\cdots,-n;-k;
\alpha+1\cdots,\alpha+1,1;\frac{1}{q},\cdots,\frac{1}{q},1\right).
\end{multline*}
Taking into account (\ref{eq:iq}), (\ref{eq:line}) and the orthogonality relation of the polynomials
$L_n^{(\alpha)}(x)$, one finally has that the term with $k=0$ is the only non-vanishing contribution
to $I_q\cor{L_n^{(\alpha)}}$, so that
\be\label{eq:iq2}
I_q\cor{L_n^{(\alpha)}}=\frac{1}{q^{\alpha q+1}}\Theta_0 \pare{\alpha q,0,2q,\{n\},\{\alpha\},\left\{\frac{1}{q}\right\}},
\ee
with
\begin{multline}
\Theta_0 \pare{\alpha q,0,2q,\{n\},\{\alpha\},\left\{\frac{1}{q}\right\}}=\Gamma(\alpha q+1)\binom{n+\alpha}{n}^{2q}\\
\times F_A^{(2 q+1)}\left(\alpha q+1;-n,\cdots,-n;0;\alpha+1,
\cdots,\alpha+1,1;\frac{1}{q},\cdots,\frac{1}{q},1\right).
\label{eq:theta0}
\end{multline}
Then, the entropic moments of the Laguerre polynomials have, according to Eqs. 
(\ref{eq:wq2}) and (\ref{eq:iq2}), the following expression
\be\label{eq:wq3}
W_q\cor{\rho_{n,\alpha}}= \cor{\frac{n!}{\Gamma(\alpha+n+1)}}^q \frac{1}{q^{\alpha q+1}}
\Theta_0 \pare{\alpha q,0,2q,\{n\},\{\alpha\},\left\{\frac{1}{q}\right\}}.
\ee

Finally, from Eqs. (\ref{eq:lq}), (\ref{eq:theta0}) and (\ref{eq:wq3}) one has that the Renyi 
entropy of order $q$ of the Laguerre polynomials is given by
\begin{eqnarray}
{\cal L}_q^R \cor{\rho_{n,\alpha}}&=& \cor{ \pare{\frac{n!}{\Gamma(\alpha+n+1)}}^q
\frac{1}{q^{\alpha q+1}} \Theta_0\pare{\alpha q;\frac{1}{q},\cdots,\frac{1}{q}}}
^{-\frac{1}{q-1}}\nonumber\\
&=& \left[ \pare{\frac{n!}{\Gamma(\alpha+n+1)}}^q
 \frac{1}{q^{\alpha q+1}} \Gamma (\alpha q+1) \binom{n+\alpha}{n}^{2q}\right.\nonumber\\
&&\times F_A^{(2 q+1)} 
\left.
\pare{
\alpha q+1,-n,\cdots,-n,0;\alpha+1,...,\alpha+1,1; \frac{1}{q},\cdots,\frac{1}{q},1}\right]^{-\frac{1}{q-1}},
\label{eq_renyi_length_lauricella}
\end{eqnarray}
for every $q>0$, $q\ne1$. Some examples follow:

\begin{itemize}
\item For $n=0$ we have that
\[
\mathcal{L}_q^R[\rho_{0,\alpha}]=\left[\frac{1}{\left(\Gamma(\alpha+1)\right)^q}\frac{\Gamma(\alpha q+1)}{q^{\alpha q+1}} F_A^{(2q+1)}\left(\alpha q+1,0,\ldots,0,0;\alpha+1,\ldots,\alpha+1,1;\frac{1}{q},\ldots,\frac{1}{q},1\right)
\right]^{-\frac{1}{q-1}}.
\]
By considering the definition of the Lauricella function in \cite{srivastava_85}, this expression can be written down as
\begin{eqnarray}
\mathcal{L}_q^R[\rho_{0,\alpha}]&=&\left[\frac{1}{\left(\Gamma(\alpha+1)\right)^q}\frac{\Gamma(\alpha q+1)}{q^{\alpha q+1}}\right.\nonumber\\
&&\left.\times \sum_{j_1,\ldots,j_{2q+1}=0}^{+\infty} \frac{(\alpha q+1)_{j_1+\cdots+j_{2q+1}}(0)_{j_1}\cdots (0)_{j_{2q+1}}}{(\alpha+1)_{j_1}\cdots(\alpha+1)_{j_{2q}}(1)_{2q+1}}
\frac{\left(\frac{1}{q}\right)^{j_1+\cdots+j_{2q}}}{j_1!\cdots j_{2q+1}!}\right]^{-\frac{1}{q-1}} \nonumber\\
&=&\cor{\frac{1}{\Gamma(\alpha+1)^q}
\frac{\Gamma(\alpha q+1)}{q^{\alpha q+1}}}^{-\frac{1}{q-1}},
\label{eq_case_n0_algebraic}
\end{eqnarray}
where we have taken into account that only the term with $j_1=\cdots=j_{2q+1}=0$ is different from zero in the previous sum.

\item For $n=1$ we have
\[
\mathcal{L}_q^R[\rho_{1,\alpha}]=\left[\frac{\Gamma(\alpha q+1)}{\left(\Gamma(\alpha+2)\right)^q}\frac{(1+\alpha)^{2q}}{q^{\alpha q+1}}  F_A^{(2q+1)}\left(\alpha q+1,-1,\ldots,-1,0;\alpha+1,\ldots,\alpha+1,1;\frac{1}{q},\ldots,\frac{1}{q},1\right)
\right]^{-\frac{1}{q-1}},
\]
that yields
\begin{eqnarray*}
\mathcal{L}_q^R[\rho_{1,\alpha}]&=&\left[
\frac{\Gamma(\alpha q+1)}{\left(\Gamma(\alpha+2)\right)^q}\frac{(1+\alpha)^{2q}}{q^{\alpha q+1}}
\right.\\
&&\left.\times \sum_{j_1,\ldots,j_{2q+1}=0}^{+\infty} \frac{(\alpha q+1)_{j_1+\cdots+j_{2q+1}}(-1)_{j_1}\cdots (-1)_{j_{2q}}(0)_{j_{2q+1}}}{(\alpha+1)_{j_1}\cdots(\alpha+1)_{j_{2q}}(1)_{2q+1}}
\frac{\left(\frac{1}{q}\right)^{j_1+\cdots+j_{2q}}}{j_1!\cdots j_{2q+1}!}\right]^{-\frac{1}{q-1}},
\end{eqnarray*}
as $(-1)_0=1$, $(-1)_1=-1$ and $(-1)_i=0, i\ge 2$, only the terms with $j_1,\ldots,j_{2q}$ equal to 0 and 1 are different from zero. Then, by considering the number of terms with a given number $k$ of indices equal to 1, this expression can be straightforwardly reduced to
\begin{equation}
\mathcal{L}_q^R[\rho_{1,\alpha}]=\left[
\frac{\Gamma(\alpha q+1)}{\left(\Gamma(\alpha+2)\right)^q}\frac{(1+\alpha)^{2q}}{q^{\alpha q+1}}
\sum_{k=0}^{2q}\binom{2q}{k}(\alpha q+1)_k \left(\frac{-1}{q(\alpha+1)}\right)^k\right]^{-\frac{1}{q-1}},
\label{eq_case_n1_algebraic}
\end{equation}
which can be further simplified as
\[
{\cal L}_q^R\cor{\rho_{1,\alpha}}=\left[\frac{\Gamma(\alpha q+1)(1+\alpha)^{2q}}{(\Gamma(\alpha+2))^q q^{\alpha q+1}}
\,_2F_0\left(
\begin{array}{c}
-2q,\alpha q+1\\
-
\end{array}
;\frac{-1}{q(\alpha+1)}
\right)
\right]^{-\frac{1}{q-1}}.
\]
\end{itemize}

\subsection{Combinatorial approach}

In this approach we begin with the explicit expression of the Laguerre polynomials given by
\[
\widetilde{L}_n^{(\alpha)} (x)=\sum_{k=0}^n c_k x^k,
\]
with
\be\label{eq:ck}
c_k^{(n,\alpha)}=\sqrt{\frac{\Gamma(n+\alpha+1)}{n!}}\frac{(-1)^k}{\Gamma(\alpha+k+1)}\binom{n}{k},
\ee

Recently (\cite{sanchezmoreno:jcam09}; see appendix) it has been 
found that an integer power of a polynomial can be
 expressed by use of the multivariate Bell polynomials of 
Combinatorics \cite{riordan_80}. This result
 applied to the Laguerre polynomials gives
\be\label{eq:laguerrebell}
\left[ \widetilde{L}_n^{(\alpha)} (x) \right]^p =\sum_{k=0}^{np} \frac{p!}{(k+p)!}
B_{k+p,p}(c_0^{(n,\alpha)},2! c_1^{(n,\alpha)},..., (k+1)! c_k^{(n,\alpha)}) x^k,
\ee
with $c_i^{(n,\alpha)}=0$ for $i>n$, and the remaining coefficients are given by Eq. (\ref{eq:ck}). Moreover, 
the Bell polynomials are given by
\begin{equation}
B_{m,l}(a_1,a_2,\ldots,a_{m-l+1})=\sum_{\hat{\pi}(m,l)}
\frac{m!}{j_1! j_2!\cdots j_{m-l+1}!}\left(\frac{a_1}{1!}\right)^{j_1} 
\left(\frac{a_2}{2!}\right)^{j_2}\cdots \left(\frac{a_{m-l+1}}{(m-l+1)!}
\right)^{j_{m-l+1}},
\label{eq_bell_poly_definition}
\end{equation}
where the sum runs over all partitions $\hat{\pi}(m,l)$ such that 
\[
j_1+j_2+\cdots+j_{m-l+1}=l, \quad \text{and}\quad j_1+2j_2+\cdots+(m-l+1)j_{m-l+1}=m.
\]

The replacement of expression (\ref{eq:laguerrebell}) with $p=2q$ into 
Eq. (\ref{eq:wq}) yields the value
\ben\label{eq:wq4}
W_q[\rho_{n,\alpha}]&=&\sum_{k=0}^{2nq} \frac{(2q)!}{(k+2q)!}B_{k+2q,2q}(c_0^{(n,\alpha)},2!c_1^{(n,\alpha)},\ldots,(k+1)!c_k^{(n,\alpha)})
\int_{0}^\infty  x^{q\alpha} e^{-qx}x^kdx=\nonumber\\
&=&\sum_{k=0}^{2nq}\frac{\Gamma(\alpha q+k+1)}{q^{\alpha q+k+1}} \frac{(2q)!}{(k+2q)!}
B_{k+2q,2q}\pare{c_0^{(n,\alpha)},2!c_1^{(n,\alpha)},...,(k+1)!c_k^{(n,\alpha)}},
\een
where the parameters $c_i^{(n,\alpha)}$ are given by Eq. (\ref{eq:ck}), keeping in mind that $c_i^{(n,\alpha)}=0$ 
for every $i>n$, so that the only non-vanishing terms correspond to those with $j_{i+1}=0$ so 
that $(c_i^{(n,\alpha)})^{j_{i+1}}=1$ for every $i>n$.

It is worthwhile to check that for $q=1$ one has that 
\[
W_1\cor{\rho_{n,\alpha}}=\int_0^\infty \rho_{n,\alpha}(x)dx=1,
\]
and that for $q=2$ we have 
\[
W_2\cor{\rho_{n,\alpha}}=\sum_{k=0}^{4n} \frac{\Gamma(2\alpha+k+1)}{2^{2\alpha+k+1}}
\frac{24}{(k+4)!}B_{k+4,4}\pare{c_0^{(n,\alpha)},2!c_1^{(n,\alpha)},...,(k+1)!c_k^{(n,\alpha)}},
\]
for the Onicescu information \cite{onicescu_craspa66} of the Laguerre polynomials. Finally, from Eqs. (\ref{eq:lq}) and (\ref{eq:wq4}) one has the following alternative expression for the $q$th-order Renyi length 
of the Laguerre polynomial for $q=2,3,\ldots$,
\begin{equation}
{\cal L}_q^R\cor{\rho_{n,\alpha}}= \left[ \sum_{k=0}^{2nq}\frac{\Gamma(\alpha q+k+1)}
{q^{\alpha q+k+1}} \frac{(2q)!}{(k+2q)!}
B_{k+2q,2q}\pare{c_0^{(n,\alpha)},2!c_1^{(n,\alpha)},...,(k+1)!c_k^{(n,\alpha)}}\right]^{-\frac{1}{q-1}},
\label{eq_renyi_length_bell}
\end{equation}
which for $q=2$ yields the value 
\[
{\cal L}_2^R\cor{\rho_{n,\alpha}}=\cor{\sum_{k=0}^{4n} \frac{\Gamma(2\alpha+k+1)}{2^{2\alpha+k+1}}
\frac{24}{(k+4)!}B_{k+4,4}\pare{c_0^{(n,\alpha)},2!c_1^{(n,\alpha)},...,(k+1)!c_k^{(n,\alpha)}}}^{-1},
\]
for the Onicescu or second-order Renyi length \cite{onicescu_craspa66} of the Laguerre polynomials.

With these expressions we obtain the same values 
of ${\cal L}_q^R[\rho_{0,\alpha}]$ and ${\cal L}_q^R[\rho_{1,\alpha}]$ as in the previous subsection:

\begin{itemize}
\item For $n=0$ we have
\[
\mathcal{L}_q^R[\rho_{0,\alpha}] =\left[\frac{\Gamma(\alpha q+1)}{q^{\alpha q+1}} B_{2q,2q}(c_0^{(0,\alpha)},0,\ldots,0)\right]^{-\frac{1}{q-1}}.
\]
Since only the coefficient $c_0^{(0,\alpha)}$ is different from zero, the previous Bell polynomial contains only one term with index $j_1=2q$ (see Eq. (\ref{eq_bell_poly_definition})). This yields
\[
\mathcal{L}_q^R[\rho_{0,\alpha}]
=\left[ \frac{\Gamma(\alpha q+1)}{q^{\alpha q+1}}\left(c_0^{(1,\alpha)}\right)^{2q}\right]^{-\frac{1}{q-1}}
=\left[\frac{1}{(\Gamma(\alpha q+1))^q}\frac{\Gamma(\alpha q+1)}{q^{\alpha q+1}}\right]^{-\frac{1}{q-1}},
\]
in accordance with the result (\ref{eq_case_n0_algebraic}) obtained with the previous method.

\item For $n=1$ we have
\[
\mathcal{L}_q^R[\rho_{1,\alpha}] =\left[\sum_{k=0}^{2q}\frac{\Gamma(\alpha q+k+1)}{q^{\alpha q+k+1}}
\frac{(2q)!}{(k+2q)!} B_{k+2q,2q}(c_0^{(1,\alpha)},2c_1^{(1,\alpha)}0,\ldots,0)\right]^{-\frac{1}{q-1}}.
\]
Since the only coefficients different from zero are $c_0^{(1,\alpha)}$ and $c_1^{(1,\alpha)}$, the previous Bell polynomials contain only one term with indices $j_1=2q-k$ and $j_2=k$ (see Eq. (\ref{eq_bell_poly_definition})). This yields
\begin{eqnarray*}
\mathcal{L}_q^R[\rho_{1,\alpha}]
&=&\left[\sum_{k=0}^{2q} \frac{\Gamma(\alpha q+k+1)}{q^{\alpha q+k+1}}\frac{(2q)!}{(2q-k)!k!} \left(c_0^{(1,\alpha)}\right)^{2q-k} \left(c_1^{(1,\alpha)}\right)^{k}\right]^{-\frac{1}{q-1}}\\
&=&\left[\frac{\Gamma(\alpha q+1)(1+\alpha)^{2q}}{(\Gamma(\alpha+2))^q q^{\alpha q+1}}
\sum_{k=0}^{2q}\binom{2q}{k}(\alpha q+1)_k \left(\frac{-1}{\alpha(q+1)}\right)^k\right]^{-\frac{1}{q-1}},
\end{eqnarray*}
in accordance with the corresponding result (\ref{eq_case_n1_algebraic}) obtained with the first method.

\end{itemize}

The equivalence of Eqs. (\ref{eq_renyi_length_lauricella}) and (\ref{eq_renyi_length_bell}) for a generic $n$ can not be easily proved because of the non-trivial special functions involved.

\section{Shannon length: Asymptotics and sharp bounds}
The goal of this Section is twofold. First, to study the asymptotics
of the Shannon spreading length $N\cor{\rho_{n,\alpha}}$
 of the orthonormal Laguerre polynomials $\widetilde{L}_{n}^{(\alpha)}$ and its relation to the standard 
deviation $\pare{\Delta x}_{n,\alpha}$. Second, to find sharp upper bounds to 
$N\cor{\rho_{n,\alpha}}$  by use of an information-theoretic optimization procedure. 

Although many results have been recently published in the literature (see e.g. \cite{igashov:itsf99,borwein:sjna08}) 
about the asymptotics of the Laguerre polynomials themselves, they have not yet been 
successfully used to obtain the asymptotics of functionals of these mathematical functions beyond 
the $L_p$-norm method of Aptekarev et al \cite{aptekarev:rassm95,dehesa:jmp98}. Here we use the 
results provided by this method to fix the asymptotics of the Shannon length of these polynomials 
and its relation to the standard deviation. From Eq. (\ref{eq:shannon}) we have that 
\be\label{eq:expshannon}
N\cor{\rho_{n,\alpha}}=\exp{\pare{S\cor{\rho_{n,\alpha}}}},
\ee
where
\[
 S\cor{\rho_{n,\alpha}}:=-\int_0^{\infty} \omega_{\alpha}(x)\cor{\widetilde{L}_n^{(\alpha)}(x)}^2
\log{\key{\omega_{\alpha}(x)\cor{\widetilde{L}_n^{(\alpha)}(x)}^2}}dx=
E_n\cor{\widetilde{L}_n^{(\alpha)}}+J_n\cor{\widetilde{L}_n^{(\alpha)}},
\]
with the following entropic functionals \cite{dehesa:jcam01,dehesa:jmp98,sanchezruiz:jcam00}
\be\label{eq:shannonsep}
E_n\cor{\widetilde{L}_n^{(\alpha)}}=-\int_0^\infty \omega_{\alpha} (x)
\cor{\widetilde{L}_n^{(\alpha)}(x)}^2 \log{\cor{\widetilde{L}_n^{(\alpha)}(x)}^2}
dx,
\ee
and
\be\label{eq:in}
J_n\cor{\widetilde{L}_n^{(\alpha)}}=-\int_0^\infty \omega_{\alpha} (x)
\cor{\widetilde{L}_n^{(\alpha)}(x)}^2 \log{\omega_{\alpha}(x)}dx=
2n+\alpha+1-\alpha\psi\pare{\alpha+n+1}.
\ee

Moreover, the use to the $L_p$-norm method of Aptekarev et al \cite{aptekarev:rassm95}
has permitted to find \cite{dehesa:jcam01}  the following values for the asymptotics of 
$E_n\cor{\widetilde{L}_n^{(\alpha)}}$
\be\label{eq:en}
E_n\cor{\widetilde{L}_n^{(\alpha)}}=-2n+\pare{\alpha+1}\log{(n)}-\alpha-2+\log{(2\pi)}+o(1).
\ee

Then, according to Eqs. (\ref{eq:shannonsep}), (\ref{eq:in}) and (\ref{eq:en}), one has that the 
asymptotical behaviour
\[
S\cor{\rho_{n,\alpha}}=(\alpha+1)\log(n)-\alpha\log(n+\alpha+1)-1+
\log(2\pi)+o(1),
\]
for the Shannon entropy, and
\be\label{eq:asympn}
N\cor{\rho_{n,\alpha}}\simeq \frac{2\pi}{e} \frac{n^{\alpha+1}}{(n+\alpha+1)^\alpha},
\ee
for the Shannon length of the orthonormal Laguerre polynomials. Where we have used the logarithmic asymptotic behaviour of the digamma function $\psi$ (see Eq. (6.3.18) from \cite{abramowitz_64}). Moreover, from Eqs. 
(\ref{eq:deltax}) and (\ref{eq:asympn}) one finds that
\be\label{eq:shannonasimp}
N\cor{\rho_{n,\alpha}}= \frac{\pi\sqrt{2}}{e}\;
\pare{\Delta x}_{n,\alpha}+O(1),
\ee
between the asymptotical values of the Shannon length and the standard deviation of 
the polynomial $\widetilde{L}_n^{(\alpha)}$. It is worth noting that this relation fulfils 
the general inequality (\ref{eq:inequalities}) which mutually relates the Shannon length 
and the standard deviation for general densities. Moreover, the relation 
(\ref{eq:shannonasimp}) for the Rakhmanov densities of Laguerre polynomials is 
also satisfied by the Rakhmanov densities of Hermite 
\cite{sanchezmoreno:jcam09,devicente_04} and Jacobi \cite{guerrero:preprint10,devicente_04} polynomials.

Let us now find sharp upper bounds to the Shannon length $N\cor{\rho_{n,\alpha}}$ 
by taking into account the non-negativity of the relative Shannon entropy (also called Kullback-Leibler 
entropy) of two arbitrary probability densities $\rho(x)$ and $f(x)$:
\[
KL[\rho,f]=\int\rho(x)\log\frac{\rho(x)}{f(x)}dx.
\]
The Jensen inequality implies that $KL[\rho,f]\ge0$. Then, as
\[
 KL[\rho,f]=\int \rho(x)\log\rho(x)dx -\int\rho(x)\log f(x)dx = -S[\rho]-\int\rho(x)\log f(x)dx,
\]
we have that the 
Shannon entropy of $\rho_{n,\alpha}(x)$ is bounded from above by means of
\begin{equation}
S\cor{\rho_{n,\alpha}}\le-\int_0^{\infty} \rho_{n,\alpha}(x) \log{f(x)}dx.
\label{eq:shannon_bound_f}
\end{equation}

This expression produces an infinite set of upper bounds to the
Shannon entropy of Laguerre polynomials. Furthermore, to obtain an expression in terms of useful expectation values like $\langle x^b\rangle$ and $\langle \log x\rangle$, the choice of $f(x)$ is
\be
f(x)=\frac{b a^{\frac{1+m}{b}}}{\Gamma\left(\frac{1+m}{b}\right)}x^m e^{-ax^b}; \; m>-1,a>0,\,b\in\mathbb{N}^+,\, 0\le x<\infty,
\label{eq:f_x_probe_density}
\ee
which is normalized to unity. Evaluating now the bound in (\ref{eq:shannon_bound_f}) we obtain
\[
-\int_0^{+\infty}\rho(x)\log\rho(x)dx=\log\frac{\Gamma\left(\frac{1+m}{b}\right)}{b a^{\frac{1+m}{b}}}
-m\langle\log x\rangle+a\langle x^b\rangle.
\]
Differentiating this expression with respect to the parameter $a$, and equating it to zero we obtain the value $a=\frac{1+m}{b\langle x^b\rangle}$, that is a minimum given the convexity of the previous expression. This yields the bound
\be\label{eq:shannoncota}
S\left[ \rho_{n,\alpha}\right] \le \log\left[\frac{\Gamma(\beta) 
e^\beta}{b\beta^\beta}\langle x^b\rangle^\beta\right]-m\langle \log x\rangle;\;b>0,\;m>-1,\;\beta=\frac{1+m}{b},
\ee
following the lines of Refs \cite{angulo:pra94,angulo:jcp92}. Then, according to Eqs. (\ref{eq:expshannon}) and 
(\ref{eq:shannoncota}), we have the following set of infinite sharp bounds
\be\label{eq:expshannonbound}
N\left[\rho_{n,\alpha}\right] \le \frac{\Gamma\left( \beta\right)  e^\beta} {b\beta^\beta} 
\esp{x^b}^\beta e^{-m\esp{\log x}};\;m>-1,\;b>0,
\ee
for the Shannon length of the Laguerre polynomial $L_n^{(\alpha)}(x)$. For 
$m=0$ we have the upper bound
\be\label{eq:expshannonboundm0}
N\left[\rho_{n,\alpha}\right] \le \frac{\Gamma\left( \frac{1}{b}\right) \left( be \right)^\frac{1}{b}} {b} \esp{x^b}^\frac{1}{b},\;b>0.
\ee
This bound is particularly interesting because it only depends on the expectation value $\esp{x^b}$; the expectation value $\langle\log x\rangle$ is, at times, unavailable or difficult to evaluate.

\section{Some computational issues}
In this section we study various computational issues of the direct spreading measures of 
Laguerre polynomials. It is worth pointing out that there is no stable numerical algorithm for 
the computation of the Renyi and Shannon lengths of these polynomials in contrast to the case
 of orthogonal polynomials on a finite interval for which an efficient algorithm based on the 
three-term recurrence relation has been recently found by Buyarov et al \cite{buyarov:sjsc04}. 
Moreover, a naive numerical evaluation of these Laguerre functionals by means of quadratures 
is not often convenient except for the lowest-order polynomials since the increasing number of 
integrable singularities spoils any attempt to achieve a reasonable accuracy for arbitrary $n$. Here 
we carry out the following numerical study. First, we examine the numerical accuracy of the
bounds (\ref{eq:expshannonbound}) and (\ref{eq:expshannonboundm0}) to the Shannon length 
of the Laguerre polynomials $L_n^{(\alpha)}(x)$, $\alpha$ fixed, for various degrees $n$ by taking 
into account the optimal values of the parameter $b$ in (\ref{eq:expshannonboundm0}) and 
the optimal values of $(b,m)$ in (\ref{eq:expshannonbound}). These optimal values have been obtained by minimizing the corresponding inequalities numerically. Second, we study the mutual 
comparison of Fisher, Shannon and Onicescu lengths and the standard deviation of $L_n^{(\alpha)}(x)$ 
for fixed $\alpha$ and various degrees $n$. Finally, we discuss the correlation of the 
Shannon length $N(L_n^{(\alpha)})$ and the standard deviation $\left(\Delta x\right)_n$ for 
various pairs $(n,\alpha)$, which allow us to find, at times, linear relations between their
 components.

\begin{figure}
\begin{tabular}{cc}
\includegraphics[scale=0.4]{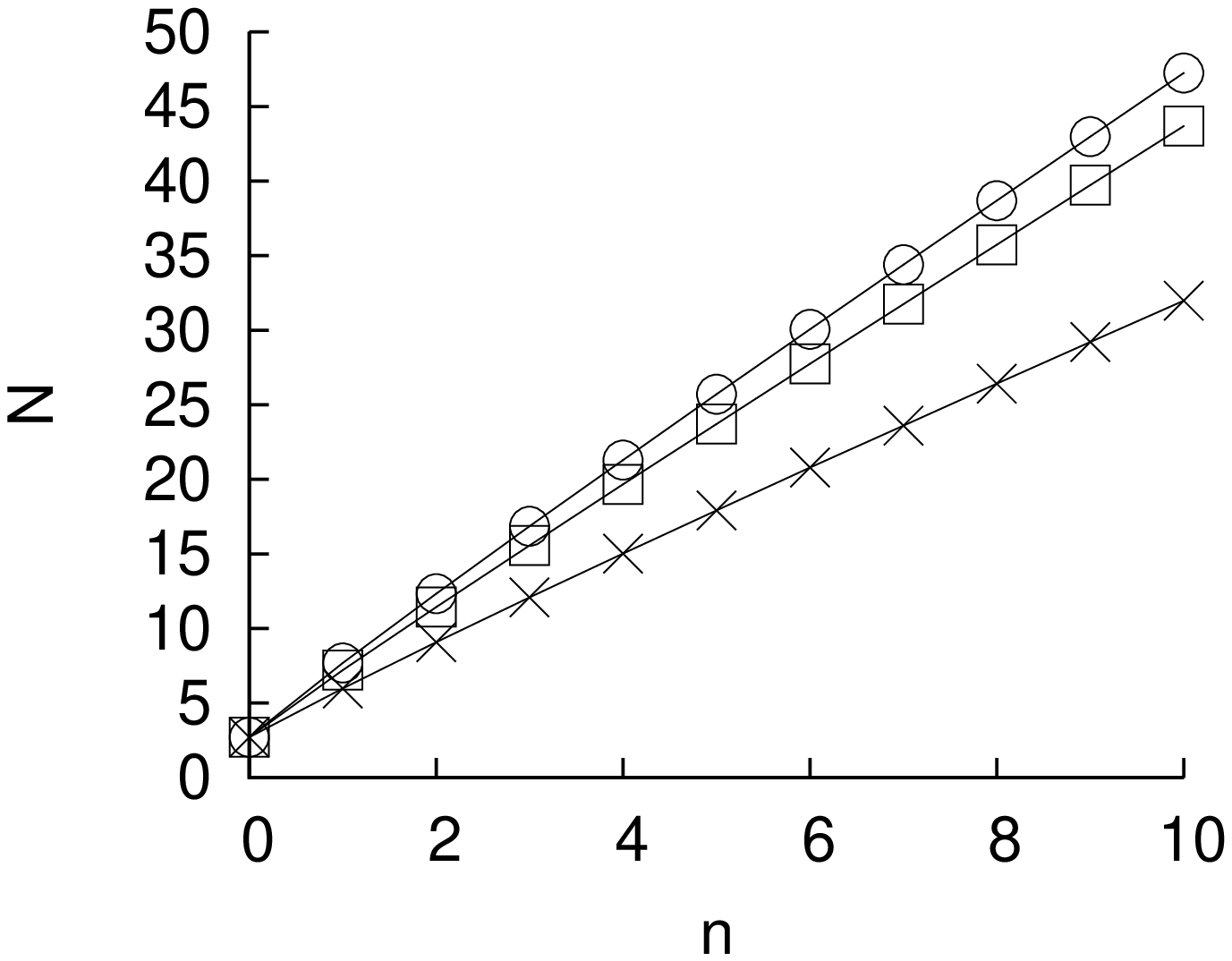}&
\includegraphics[scale=0.4]{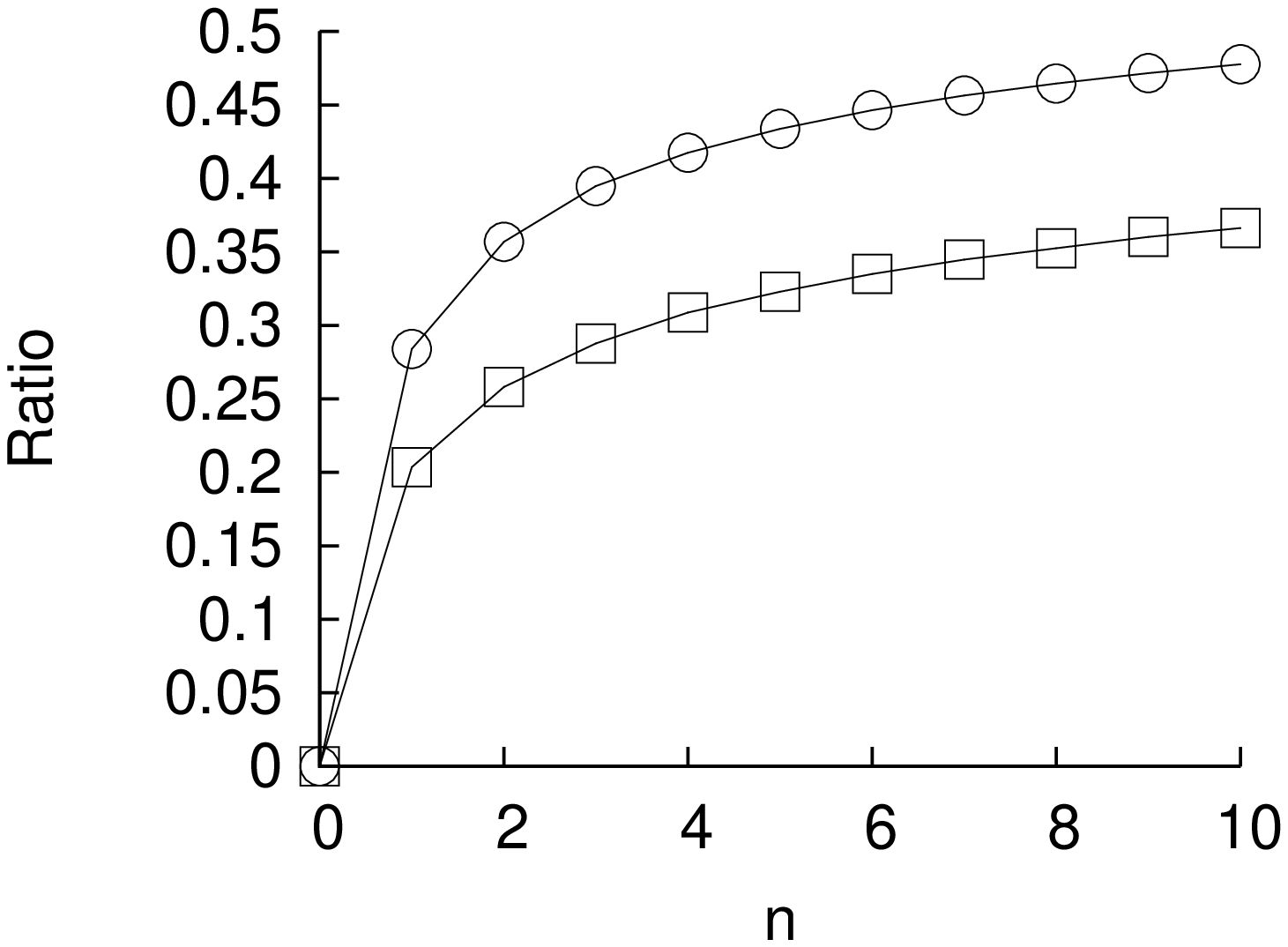}\\
\end{tabular}
\caption{Left: Shannon length ($\times$), upper bound with $m=0$ and $b=b_{\rm opt}$ ($\odot$), and upper bound with $m=m_{\rm opt}$ and $b=b_{\rm opt}$ ($\boxdot$) of the Laguerre polynomials $L_n^{(0)}(x)$, as a function of the degree $n$. Right: Relative ratios of the bounds with $m=0$ and $b=b_{\rm opt}$ ($\odot$), and with $m=m_{\rm opt}$ and $b=b_{\rm opt}$ ($\boxdot$), as a function of $n$.}
\label{fig1}
\end{figure}

In Figure \ref{fig1} it is numerically studied the accuracy of the bounds
(\ref{eq:expshannonbound}) and (\ref{eq:expshannonboundm0}) to the Shannon length of the 
Laguerre polynomial $L_n^{(\alpha=0)}(x)$ given by the optimal values $(b_{\rm opt},0)$ in 
(\ref{eq:expshannonboundm0}). This is done by comparing the corresponding optimal bounds 
with the ``numerically exact'' value of the lengths $N\left[\rho_{n,0}\right]$ for the polynomials 
with degree $n$ from $0$ to $10$. The graph on the right of the figure gives the relative ratio 
of the bound given by Eq. (\ref{eq:expshannonboundm0}) with $b=b_{\rm opt}$, and the ratio of the bound given by Eq. (\ref{eq:expshannonbound}) with $b=b_{\rm opt}$ and $m=m_{\rm opt}$. Notice that the latter bound
is always 
better than the former, as we have the parameter $m$ to adjust. The 
values of optimal pairs $(b_{\rm opt},0)$ and $(b_{\rm opt},m_{\rm opt})$ are shown in Tables \ref{tab1} and \ref{tab2}, respectively. Note that the optimum value $b_{\rm opt}$ is different when considering the bound (\ref{eq:expshannonbound}) or (\ref{eq:expshannonboundm0}). Also notice that the optimum values for $n=0$ are $(b_{\rm opt},m_{\rm opt})=(1,0)$, where the density $f(x)$ defined in (\ref{eq:f_x_probe_density}) equals the Rakhmanov density for $n=0$.
Remark that the best bounds are obtained for expectation values 
$\esp{x^b}$ where $b=b_{\rm opt}$ is an increasing function of the degree $n$ of the 
polynomial in both cases;
this is directly connected with the larger spreading of the polynomial when its degree has 
higher values.

\begin{table}
\begin{tabular}{|c|c|c|c|c|c|c|c|c|c|c|c|}
\hline
{$n$} & 0 & 1 & 2 & 3 & 4 & 5 & 6 & 7 & 8 & 9  & 10\\
\hline
{$b_{\rm opt}$} & 1 & 3 & 4 & 6 & 7 & 8 & 9 & 10 & 11 & 12 & 13\\
\hline
\end{tabular}
\caption{Values $b_{\rm opt}$ of the parameter $b$ which yield the best (i.e. lowest) upper bounds (\ref{eq:expshannonboundm0}) to the Laguerre polynomial $L_n^{(0)}(x)$ for various degrees $n$.}
\label{tab1}
\end{table}

\begin{table}
\begin{tabular}{|c|c|c|c|c|c|c|c|c|c|c|c|}
\hline
{$n$} & 0 & 1 & 2 & 3 & 4 & 5 & 6 & 7 & 8 & 9 & 10\\
\hline
{$b_{\rm opt}$} & 1 & 4 & 6 & 7 & 9 & 10 & 11 & 12 & 14 & 15& 16\\
\hline
{$m_{\rm opt}$} & 0 & -0.332 & -0.338 & -0.322 & -0.332 & -0.327 & -0.324 & -0.321 & -0.322 & -0.320 & -0.319\\
\hline
\end{tabular}
\caption{Values $(b_{\rm opt},m_{\rm opt})$ of the parameters $(b,m)$ which yield the best (i.e. lowest) upper bound (\ref{eq:expshannonbound}) to the Laguerre polynomial $L_n^{(0)}(x)$ for various degrees $n$.}
\label{tab2}
\end{table}

\begin{figure}
\begin{tabular}{cc}
\includegraphics[scale=0.4]{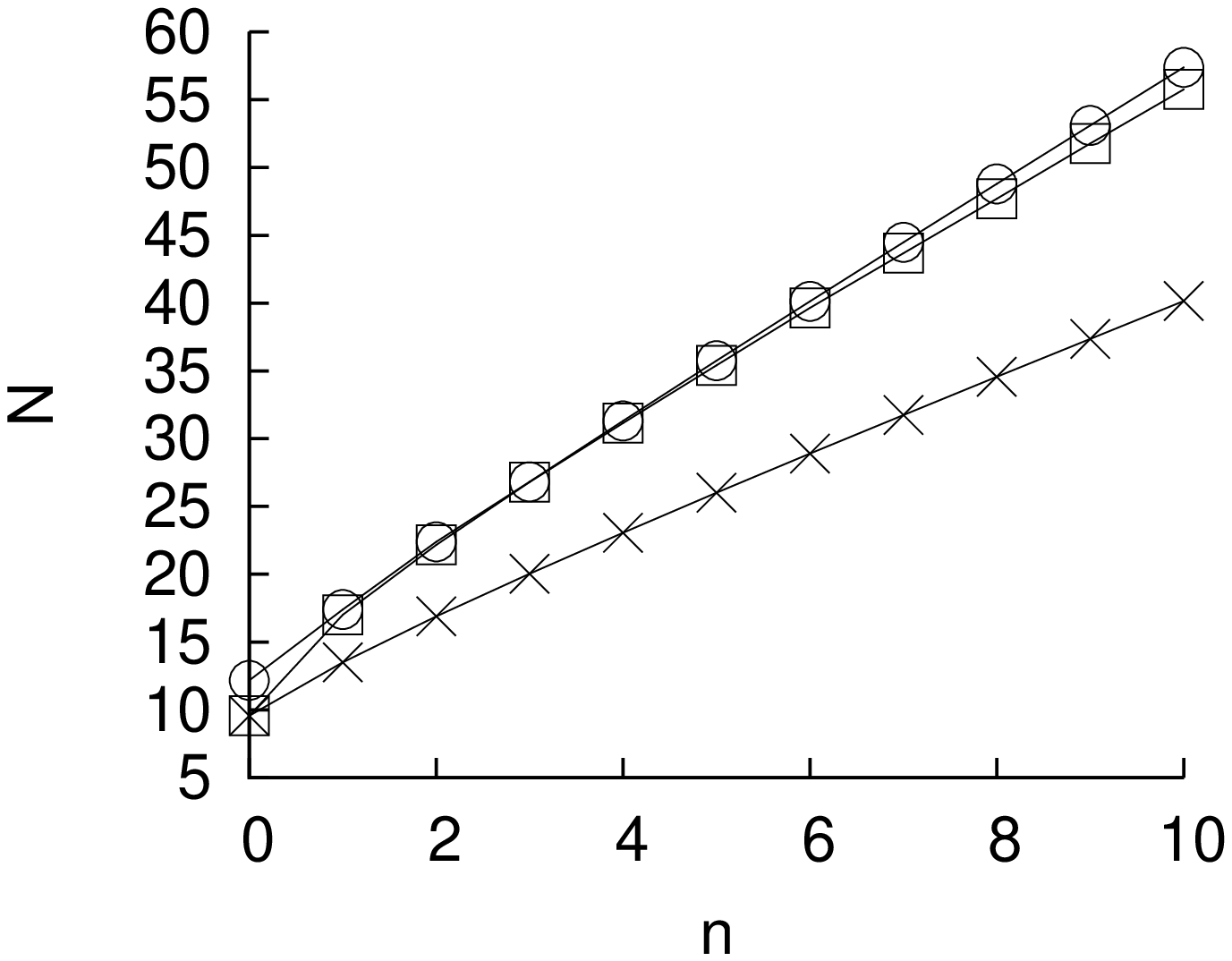}&
\includegraphics[scale=0.4]{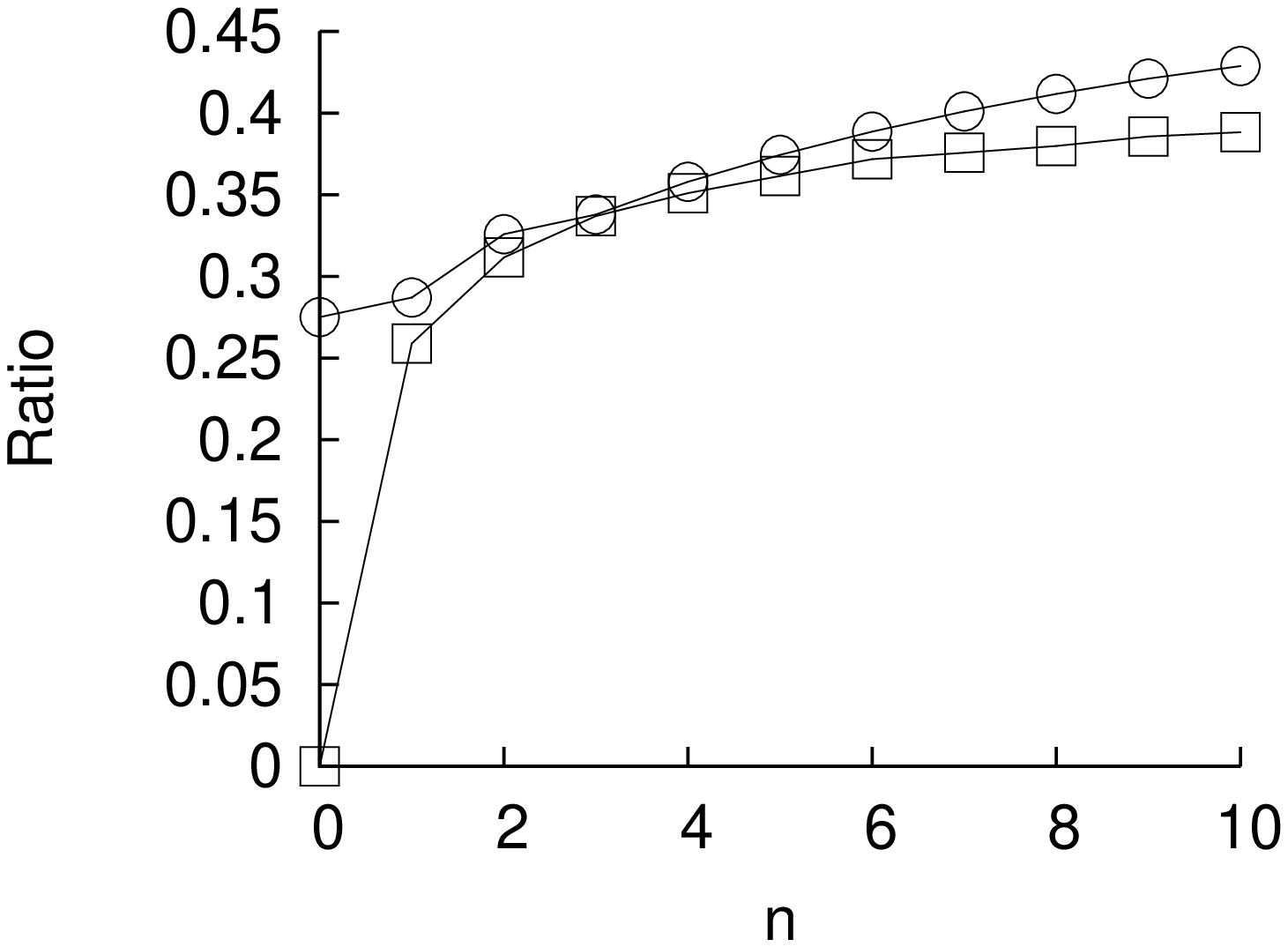}\\
\end{tabular}
\caption{Left: Shannon length ($\times$), upper bound with $m=0$ and $b=b_{\rm opt}$ ($\odot$), and upper bound with $m=m_{\rm opt}$ and $b=b_{\rm opt}$ ($\boxdot$) of the Laguerre polynomials $L_n^{(5)}(x)$, as a function of the degree $n$. Right: Relative ratios of the bounds with $m=0$ and $b=b_{\rm opt}$ ($\odot$), and with $m=m_{\rm opt}$ and $b=b_{\rm opt}$ ($\boxdot$), as a function of $n$.}
\label{fig2}
\end{figure}

To study the behaviour of the accuracy of the two previous bounds with respect to $\alpha$, we 
have done in Figure \ref{fig2} a study of the Shannon lengths $N\left[\rho_{n,5} \right]$
similar to that done in Figure \ref{fig1} for $ L_n^{(0)}(x)$. The corresponding values $(b_{\rm opt},0)$ and 
$(b_{\rm opt},m_{\rm opt})$ are given in Tables \ref{tab3} and \ref{tab4} respectively. The two graphs of 
the figure show qualitatively similar and quantitatively better results than those found in Figure \ref{fig1}.

\begin{table}
\begin{tabular}{|c|c|c|c|c|c|c|c|c|c|c|c|}
\hline
{$n$} & 0 & 1 & 2 & 3 & 4 & 5 & 6 & 7 & 8 & 9 & 10\\
\hline
{$b_{\rm opt}$} & 5 & 6 & 7 & 8 & 10 & 11 & 12 & 13 & 14 & 15 & 16\\
\hline
\end{tabular}
\caption{Values $b_{\rm opt}$  of the parameter $b$ which yield the best (i.e. lowest) upper bounds (\ref{eq:expshannonboundm0}) to the Laguerre polynomial $L_n^{(5)}(x)$ for various degrees $n$.}
\label{tab3}
\end{table}

\begin{table}
\begin{tabular}{|c|c|c|c|c|c|c|c|c|c|c|c|}
\hline
{$n$} & 0 & 1 & 2 & 3 & 4 & 5 & 6 & 7 & 8 & 9 & 10\\
\hline
{$b_{\rm opt}$} & 1 & 5  & 7 & 9 & 10 & 11 & 13 & 14 & 15 & 16 & 17\\
\hline
{$m_{\rm opt}$} & 5 & 0.288 & 0.053 & -0.049 & -0.098 & -0.131 & -0.160 & -0.177 & -0.190 & -0.201 & -0.210\\
\hline
\end{tabular}
\caption{Values $(b_{\rm opt},m_{\rm opt})$ of the parameters $(b,m)$ which yield the best (i.e. lowest) upper bound (\ref{eq:expshannonbound}) to the Laguerre polynomial $L_n^{(5)}(x)$ for various degrees $n$.}
\label{tab4}
\end{table}

\begin{figure}
\includegraphics[width=10cm]{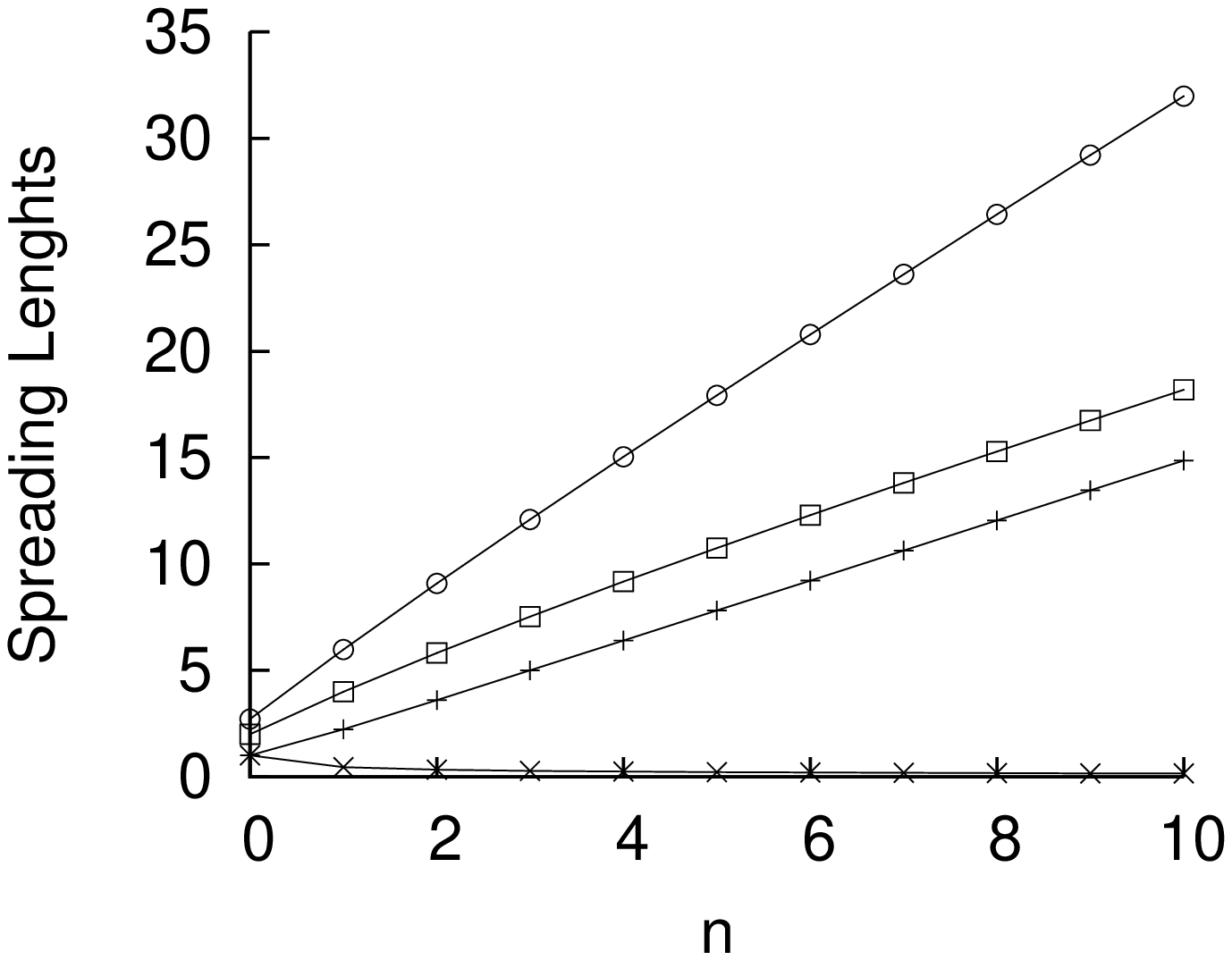}
\caption{Standard deviation $\Delta x$ ($+$), Fisher length $\delta x$ ($\times$), Onicescu length $\mathcal{L}_2$ ($\boxdot$), and Shannon length $N$ ($\odot$) of the Laguerre polynomial $L_n^{(0)}(x)$ as a function of $n$.}
\label{fig3}
\end{figure}

\begin{figure}
\includegraphics[width=10cm]{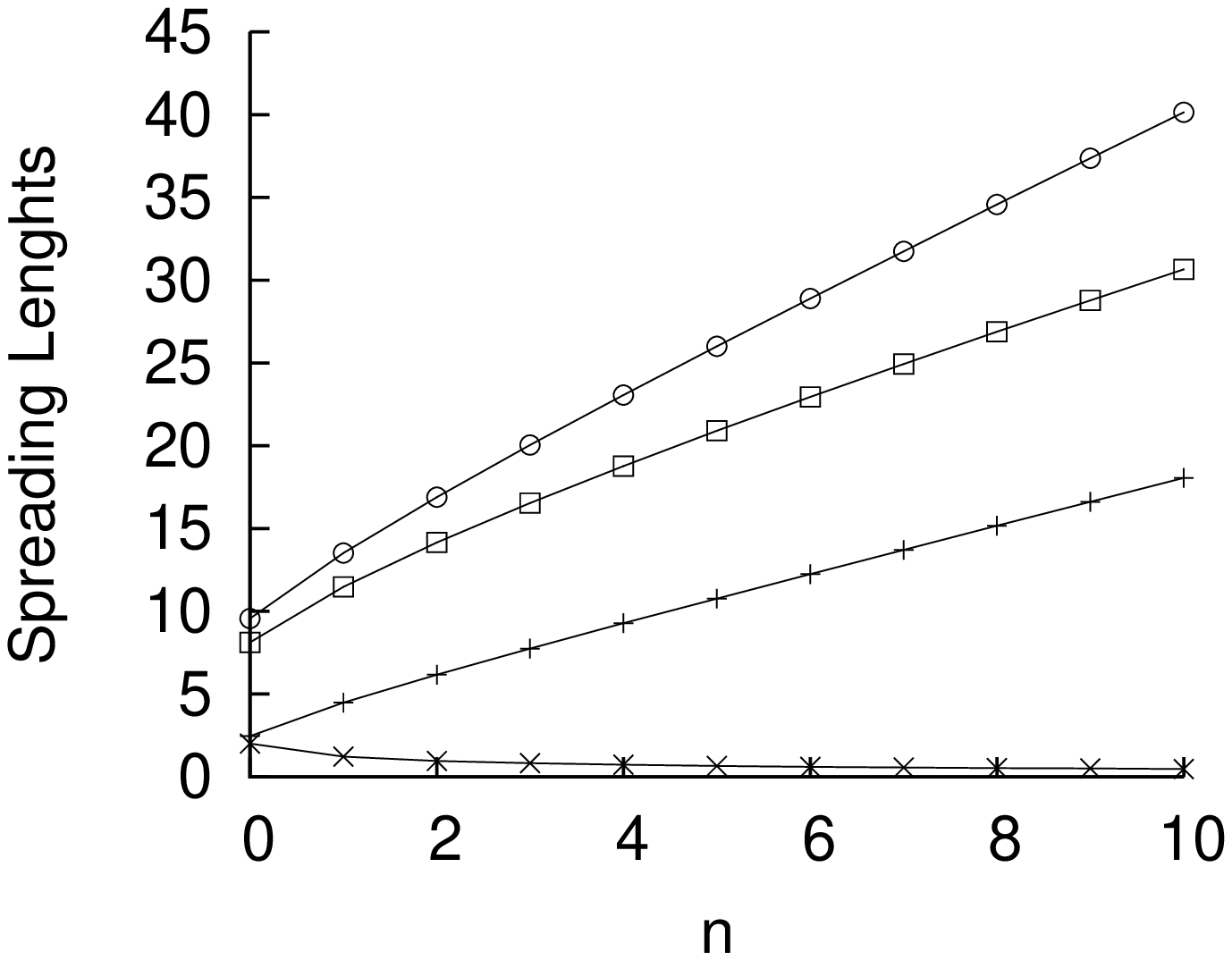}
\caption{Standard deviation $\Delta x$ ($+$), Fisher length $\delta x$ ($\times$), Onicescu length $\mathcal{L}_2$ ($\boxdot$), and Shannon length $N$ ($\odot$) of the Laguerre polynomial $L_n^{(5)}(x)$ as a function of $n$.}
\label{fig4}
\end{figure}

In Figures \ref{fig3} and \ref{fig4} we study the mutual comparison of various direct spreading 
measures (namely, the standard deviation $\Delta x$ and the Fisher, Shannon and the 
Onicescu or second-order Renyi lengths) of the Laguerre polynomials $ L_n^{(0)}(x)$ and 
$ L_n^{(5)}(x)$, respectively, when the degree $n$ varies from $0$ to $10$. Several observations 
are in order. First, all the measures with global character (standard deviation, Shannon and Renyi 
lengths) grow linearly or quasilinearly when the degree of the polynomial is increasing; essentially 
because the polynomial spreads more and more. Moreover, they
behave so that $\Delta x<{\cal L}_2<N$. Second, the (local) Fisher length decreases when the 
degree $n$ is increasing; essentially, because the polynomial becomes more and more oscillatory, 
so growing its gradient content. Third, the Fisher length has always a value smaller than all the global 
spreading measures. 

\begin{figure}
\includegraphics[width=10cm]{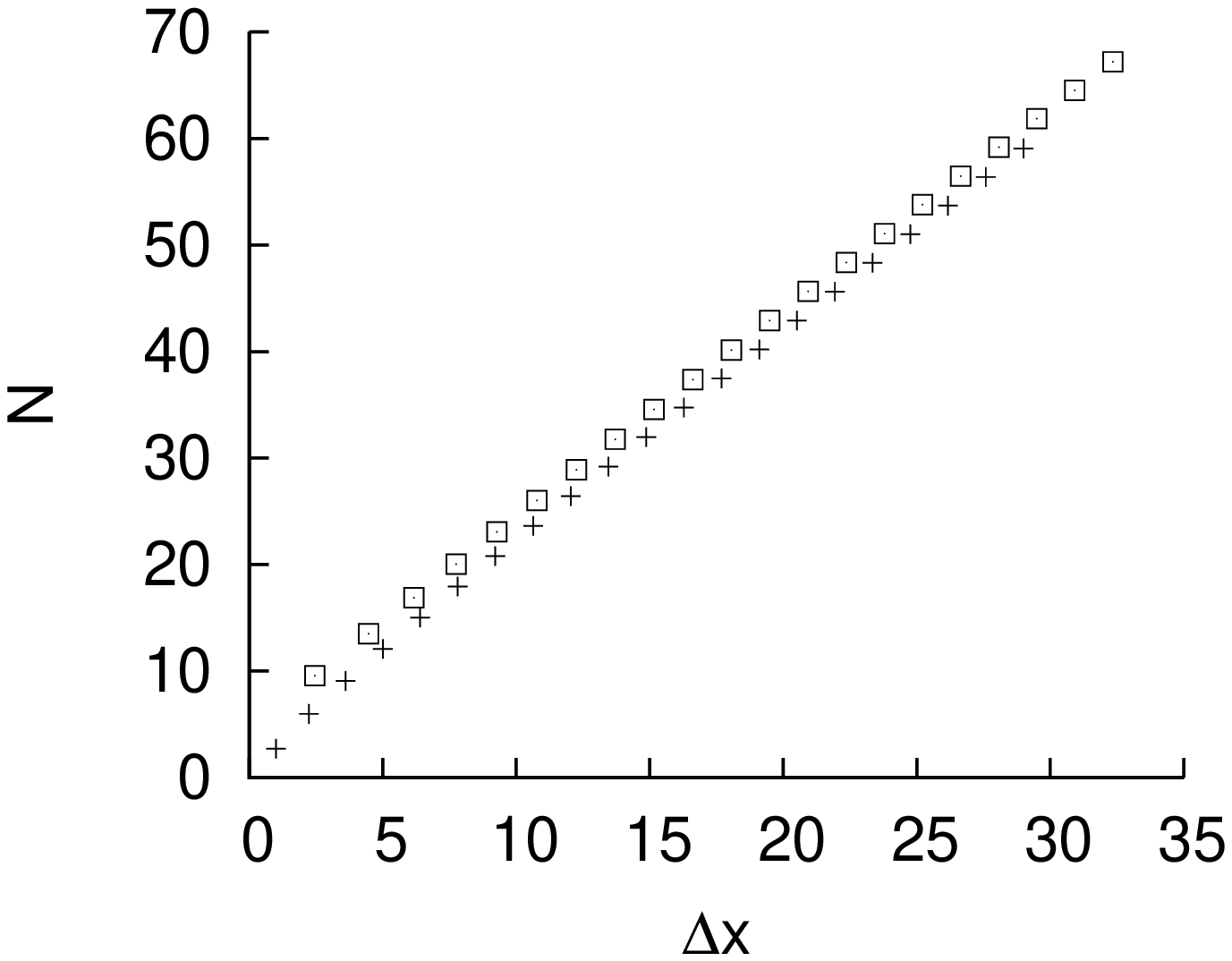}
\caption{Shannon length $N$ as a function of the standard deviation $\Delta x$ for the Laguerre polynomials $L_n^{(0)}(x)$ ($+$) and $L_n^{(5)}(x)$ ($\boxdot$), when the degree $n$ varies from 0 to 20.}
\label{fig5}
\end{figure}

Finally, in Figure \ref{fig5} we have numerically studied the connection of the Shannon length 
$N\left[ \rho_{n,\alpha} \right]$ and the standard deviation $(\Delta x)_{n,\alpha}$ of the Laguerre polynomials 
$L_n^{(\alpha)}(x)$, with $\alpha=0$ and $5$, when the degree $n$ varies from $0$ to $20$. 
This apparent quasilinear 
behaviour of the Shannon length with respect to the standard deviation is in accordance to the rigorous expression (\ref{eq:shannonasimp}), 
i.e. $N\cor{\rho_{n,\alpha}}\simeq \frac{\pi\sqrt{2}}{e} \pare{\Delta x}_{n,\alpha}$ for $n>>1$.

\section{Conclusions and open problems}
The global (standard deviation and the Renyi and 
Shannon lengths) and local (Fisher length) direct spreading measures of the Laguerre 
polynomials $L_n^{(\alpha)}(x)$ are analytically and numerically studied. Beyond the ordinary moments 
$\esp{x^k},\;k\in\mathbb{Z}$, and the standard deviation $\left( \Delta x \right)_n$, 
which have been explicitly given in terms of $(n,\alpha)$, we have developed two theoretical approaches 
of algebraic and combinatorial types to obtain two equivalent analytical expressions 
for the Renyi lengths of half-integer order. For the Shannon length, 
whose explicit value is not yet known (in fact, its calculation is a formidable task!), 
we have found sharp bounds in terms of the expectation value $\esp{x^k}$ and/or the logarithmic
expectation value $\esp{\log x}$ by means of an information-theoretic-based optimization 
procedure.

Moreover, the linear correlation of the Shannon length and the standard deviation 
for the Laguerre polynomials $L_n^{(\alpha)}(x)$ with large degree $n$ is underlined. 
In fact, the correlation factor is not only independent on the parameter $\alpha$ but, most 
importantly, it is the same as for the remaining hypergeometric families on a finite 
interval (Jacobi polynomials) \cite{guerrero:preprint10,devicente_04} or on the whole real line (Hermite polynomials) \cite{sanchezmoreno:jcam09,devicente_04}.

Then we carried out a numerical study of the four direct spreading measures of 
Laguerre polynomials. Let us remark, among other results, that the Fisher length has 
the smallest value, and the Shannon length depends quasilinearly on the standard deviation. 

Finally, let us highlight a number of open information-theoretic problems related to 
Laguerre polynomials: (i) to find the asymptotics of the entropic moments and, 
subsequently, the Renyi lengths in the spirit of \cite{aptekarev:rassm95,dehesa:jmp98}, (ii) to identify the most general class of polynomials 
for which the asymptotical relation (\ref{eq:shannonasimp}) of the Shannon length and 
the standard deviation is fulfilled, and (iii) to characterize the most general class 
of polynomials for which the ratio between these two direct spreading measures is a constant 
(i.e., it does not depend on the degree nor the parameters of the polynomials) as 
already pointed out in \cite{devicente_04}.

\section{Acknowledgments}
We are very grateful to Junta de Andaluc\'{\i}a for the grants FQM-2445 and FQM-4643, and the Ministerio de Ciencia e 
Innovaci\'on for the grant FIS2008-02380. We belong to the research group FQM-207. Daniel Manzano
 acknowledges the fellowship BES-2006-13234.

\bibliography{spreading_laguerre}
\bibliographystyle{elsarticle-num}

\end{document}